\documentclass[singlecolumn,superscriptaddress,showpacs,preprintnumbers,
               amsmath,amssymb]{revtex4}

\usepackage{float}
\usepackage{graphicx}   
\usepackage{dcolumn}    
\usepackage{bm}         
\usepackage{color}      
\usepackage{booktabs}
\usepackage[caption=false]{subfig}
\usepackage{graphicx}  
\usepackage{dcolumn}   
\usepackage{bm}        
\usepackage{verbatim}   
\makeatletter           
\newcommand*{\rom}[1]{\expandafter\@slowromancap\romannumeral #1@}
\makeatother              
\usepackage{textcomp}     
\usepackage{textgreek}
\definecolor{renk}{RGB}{255,0,0}

\begin{document}


\title{Simultaneous Production of Lepton Pairs at Ultra-peripheral Relativistic Heavy Ion Collisions}
\author{E.~Kurban}
\email{erknkurban@gmail.com}
\author{M.~C.~G\"{u}\c{c}l\"{u}}
\email{guclu@itu.edu.tr}
\affiliation{
	Department of Physics,Istanbul Technical University, Maslak-Istanbul,Turkey
}

\date{\today}

\begin{abstract}
We calculate the total cross sections and probabilities of electromagnetic productions of electron, muon and tauon pair productions simultaneously.
At the Large Hadron Collider (LHC), the available electromagnetic energy is sufficient to produce all kind of leptons coherently.
The masses of muon and tauon are large, so their Compton wavelengths are small enough to interact with the colliding nuclei. 
Therefore, the realistic nuclear form factors are included in the calculations of electromagnetic pair productions. The cross section calculations show that, at the LHC energies,
the probabilities of simultaneous productions of all kind of leptons are increased significantly compare to the RHIC energies.
Experimentally, observing this simultaneous production can give us important informations about the strong QED.

\end{abstract}

\pacs{25.75.-q; 25.75.Dw; 25.20.Lj}    
\keywords{Heavy Lepton Pair Production, QED, Nucleon Form Factors}
\maketitle

\section{\label{sec:level1}Introduction}\label{s1}
Peripheral collisions of heavy-ions have been studied for a long time \cite{bs,bb,gwuse,belks,iss,bes,bkn,bkn1} since these types of collisions contain interesting
physics problems such as lepton pair, coherent or incoherent vector-meson and some other exotic particles productions.
Very strong electromagnetic fields are produced near the peripheral heavy-ion collisions at the relativistic velocities. The electromagnetic 
fields are proportional to Lorentz factor $\gamma$, beam kinetic energy per nucleon and the charge of the ion Z. This electromagnetic field contains 
many virtual photons which are the source of lepton pair production and different types of photo-nuclear interactions.

An exact solution of the time-dependent Dirac equation for ionization and the pair production which is induced by
ultra-relativistic heavy ion collisions was done in Ref. \cite{baltz}.
This exact method could be used for other applications related to the electromagnetic lepton pairs in the future.  
The perturbative  cross section  for  free electron-positron  pair  production
at the RHIC is about 30 000 b.   On the other hand, since for the large number of energy and angular momentum states are coupled, 
non-perturbative approach for this process have difficulties, however, the method in Ref. \cite{baltz} seems to make the problem solvable. 

The Relativistic Heavy Ion Collider (RHIC) and the Large Hadron Collider (LHC) are designed to collide the fully ionized heavy ions 
in the center of mass frame with 100 GeV and 3400 GeV energies per nucleon, respectively. There are many calculations have been done 
for the RHIC energies for the production of electromagnetic lepton pair production. Nowadays we have also the LHC and its available energy is 
much larger than the RHIC. The parameters which are related with the RHIC and the LHC are listed in Table \ref{table1} and Table \ref{table2}.
The critical electric field  to produce a lepton pair and the maximum electric field for one of the heavy ions can be given as
\begin{eqnarray}
E_{crit}\approx \dfrac{(m c^{2})^{2}}{e \hbar c }, \,\,\,\,\,\,\,\,\, E_{max}\approx \dfrac{Z e \gamma}{b^{2}}.
\end{eqnarray}
Therefore, lepton pair production especially heavy lepton pair production cross section becomes quite large
and it can not be ignored.

In the past, several calculations have been done for the single pair production of leptons
\begin{eqnarray}
Z + Z \rightarrow Z + Z +l^{+}l^{-},\,\,\,\, l = e, \mu, \tau .
\end{eqnarray}
Since the mass of an electron (0.511 MeV) is much smaller than
a muon (105.66 MeV) and a tauon (1784 MeV), and the Compton wavelength of an electron (386 fm) is much larger than a muon (1.86 fm) and a tauon
(0.11 fm), the cross sections of producing electron pairs are much larger than the heavy leptons. In addition
to this, since the Compton wavelengths of muons and tauons are smaller than the radius of the colliding heavy ions (Au, Pb), 
nucleon \cite{belks,melek} and nucleus form factors are not negligible. Therefore, the realistic charge form factors play important roles 
for calculating the cross sections of the heavy lepton pair productions.

In this work, we have calculated the cross section of simultaneous production of all three lepton pairs together
\begin{eqnarray}
Z + Z \rightarrow Z + Z +e^{+}e^{-}+\mu^{+}\mu^{-}+\tau^{+}\tau^{-}.
\end{eqnarray}
The main motivation for this calculation is that the available energy in this collisions is sufficiently enough to
create all these lepton pairs simultaneously. 
Especially at the LHC energies, there are interesting processes can occur and these processes can help us to understand the strong QED.
One of them is to observe different physical processes simultaneously. For example, the STAR Collaboration measured 
the electron-positron pairs \cite{star} together with the electromagnetic excitation of both ions, predominantly to the giant dipole 
resonance. The STAR Collaboration used gold atoms at $\sqrt{s}=200$ GeV/nucleon energies. The decay of the excited nucleus 
generally emits one or two neutrons and these neutrons are detected in the forward zero degree calorimeter.

For sufficiently high energies, the production of the cross section for a single pair becomes large.
At the LHC energies, the probabilities for producing multiple pairs of leptons become important contribution to
the measured cross sections. In addition to this, the perturbative result for single-pair production violates unitary
 on the cross section for large Z and high energies. Multiple-pairs production processes solve
this single-pair production problem. Through the Poisson expression, the multiple-pairs productions cross
sections can be obtained. Experimentally \cite{vane}, only upper limits have been achieved for the measurements 
of multiple-pairs production so far.

In ultra-relativistic heavy ion collisions, there are two dominant processes that restrict the luminosity 
of the ion beams. One of them is bound-free electron-positron pair production (BFPP) that occurs with the 
capture of the electron after the free electron-positron pair production and the positron goes his way 
freely. In bound-free pair production process, produced particle especially is captured 1s ground state 
of one of the ions. The captured electron by one of the colliding ions leads to a change in the charge and 
mass of the ions and causes the ion to fall out of the beam. In this case, the ion that captures the 
particle exits from the beam. This process leads to the beam depletion. The calculation of the electron 
capture process is important for the lifetime of the beam. Recent calculation shows that electron
capture and Giant Dipole Resonance play important role for the beam luminosity \cite{ksr, star15}.

At the collisions of relativistic heavy ions, leptons are not only produced through the electromagnetic process,
in addition to this, leptons are also produced by the hadronic interactions. These leptons which are produced by the
Drell Yan process can be a possible signal for the quark-gluon plasma. However, the electromagnetically produced
lepton-pair cross section, especially electron-positron pairs, is very large so that they can mask the leptons 
originating from the quark-gluon plasma.
Therefore, it is very important to know the electromagnetic lepton pair production cross section in order to 
separate these two processes for minimizing the confusion.
\section{\label{sec:level1}Formalism}\label{s1}
\begin{figure}
	\includegraphics[width=4.95cm,height=6.56cm]{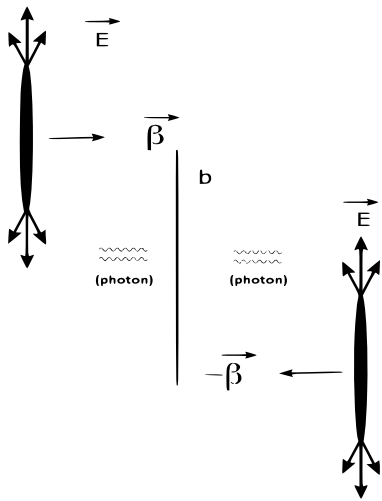} 
	\caption{Relativistic collision of two heavy ion is illustrated as schematic diagram. Impact parameter 
		between the ions is $b$, and the Lorentz contracted electric field is also shown. }
	\label{fig1} 
\end{figure}

Fig. \ref{fig1} shows the collisions of the ions where $\beta$ and -$\beta$ represent the velocities of two heavy-ions nucleus 1 and nucleus 2, respectively. The velocities are parallel to the z axis and the distance line shows impact parameter between the center of nucleus 1 and the center of nucleus 2.

The semi-classical coupling of leptons to the electromagnetic field is represented by Lagrangian density which only depends on classical four-vector potential $A^{\mu}$ and also conserves lepton number
\begin{equation}
L_{int}(x)=-\bar{\psi}\gamma_{\mu}\psi(x)A^{\mu}(x)
\end{equation}
where we can write the $A^{\mu}$ as the sum of two heavy ions
\begin{equation}
A^{\mu}(q;b)=A^{\mu}(1;q;b)+A^{\mu}(2;q;b)
\end{equation}
We can also write the components of the potential from both nuclei in momentum space as 
\begin{equation}
A^{(1,2)}_{0}(q;b)=-8\pi^{2}Z\delta(q_{0}\mp\beta.q_{z})f_{z}(q^{2})\dfrac{exp(\pm iq_{\perp}b/2)}{q_{z}^{2}+\gamma^{2}q_{\perp}^{2}}
\end{equation}
\begin{equation}
A_{x}^{(1,2)}(q)=0
\end{equation}
\begin{equation}
A_{y}^{(1,2)}(q)=0
\end{equation}
\begin{equation}
A_{z}^{(1,2)}(q)=\pm \beta.A_{0}^{(1,2)}(q)
\end{equation}
where $f_{z}(q^{2})$ is the form factor of the nucleus.
These form factors play important role for the heavy lepton pair productions. We have used 
two parameter Fermi (2pF) function (or Wood-Saxon function) for the charge distribution of the protons:
\begin{eqnarray} \label{ws}
\rho(r) = \dfrac{\rho_{0}}{1 + exp((r-R)/R_{0})}
\end{eqnarray}
\begin{table}[ht]
	\caption{Table shows the critical energies and electric fields that can produce the leptons. $\lambda_{c}$ is the Compton wavelength of the corresponding lepton. } 
	\centering  
	\begin{tabular}{c c c c} 
		\hline\hline   \\                     
		Lepton\,\,\,\,\,\,  & \,\,\,$\hslash$$\omega_{crit}$ = $2mc^{2}$ (MeV)\,\,\, &\,\,\,\, $\lambda_{c}$ (fm) \,\,\,\,& \,\,\,\,$E_{crit}$ (V/m)\,\,\, \\ [0.5ex] 
		\hline  \\                
		$e^{\pm}$ & 1.02 & 386.2 & 1.3 $\times 10^{18}$  \\ 
		$\mu^{\pm}$ & 211.4 & 1.87 & 5.7 $\times 10^{22}$  \\
		$\tau^{\pm}$ & 3568 & 0.11  & 1.6 $\times 10^{25}$  \\ [1ex]      
		\hline  
	\end{tabular}
	\label{table1} 
\end{table}
\begin{table}[ht]
	\caption{The first line in the table shows the maximum electric field of the colliding heavy ions at RHIC and LHC.
		Other lines are the ratio of between maximum electric field and the critical electric field of producing corresponding lepton.} 
	\centering  
	\begin{tabular}{c c c} 
		\hline\hline \\                        
		Parameters\,\,\,\,\,\,  & \,\,\, RHIC \,\,\, &\,\,\,\, LHC \,\,\,\, \\ [0.5ex] 
		\hline  \\                
		$E_{max}$ (V/m) & $\sim 2 \times10^{23}$ & $\sim 7 \times 10^{24}$  \\ 
		$E_{max}/E_{crit}(e^{\pm}) $ & 1.5 $\times 10^{5}$ & 5.3 $\times 10^{6}$  \\
		$E_{max}/E_{crit}(\mu^{\pm}) $ & 3.5  & 123.7   \\
		$E_{max}/{E_{crit}}(\tau^{\pm}) $ & 0.01  & 0.4   \\ [1ex]      
		\hline  
	\end{tabular}
    \label{table2}
\end{table}
where $ R $ is the radius of the nucleus and $R_{0}$ is the skin depth or diffuseness parameter. These parameters are obtained by fits to
the electron scattering data \cite{barrett} and $\rho_{0}$ is written by the normalization condition.
For symmetric nuclei, the nuclear density for a nucleus that has mass number $A$, a distance $r$
from its center is modelled in literature with a Woods-Saxon distribution as in Eq.~\ref{ws}
where $\rho_{0}=\frac{0.1694}{A}fm^{-3}$ for Au nucleus and $\rho_{0}=\frac{0.1604}{A}fm^{-3}$ for Pb nucleus. 
The radii of the gold and lead nucleus are equal to $R_{Au} = 6.38$ fm and  $R_{Pb} = 6.62$ fm respectively.
In order to see 
the effects of the proton distributions in the nucleus, we need to have an analytical expression of the Fourier
transforms of the Wood-Saxon distribution :
\begin{eqnarray}
f_{Z}(q^{2}) & = & \int^{\infty}_{0}\frac{4\:\pi}{q}\rho(r)Sin(qr)dr 
\label{ffactor}
\end{eqnarray}

where $f_{Z}(q^{2})$ is plotted in Fig. \ref{fig3}. It is clearly seen that the analytic expression and the numerical calculation
are nearly identical.
\begin{figure}[h]
  \centering
  \subfloat[]{\includegraphics[width=6.0cm,height=4.0cm]{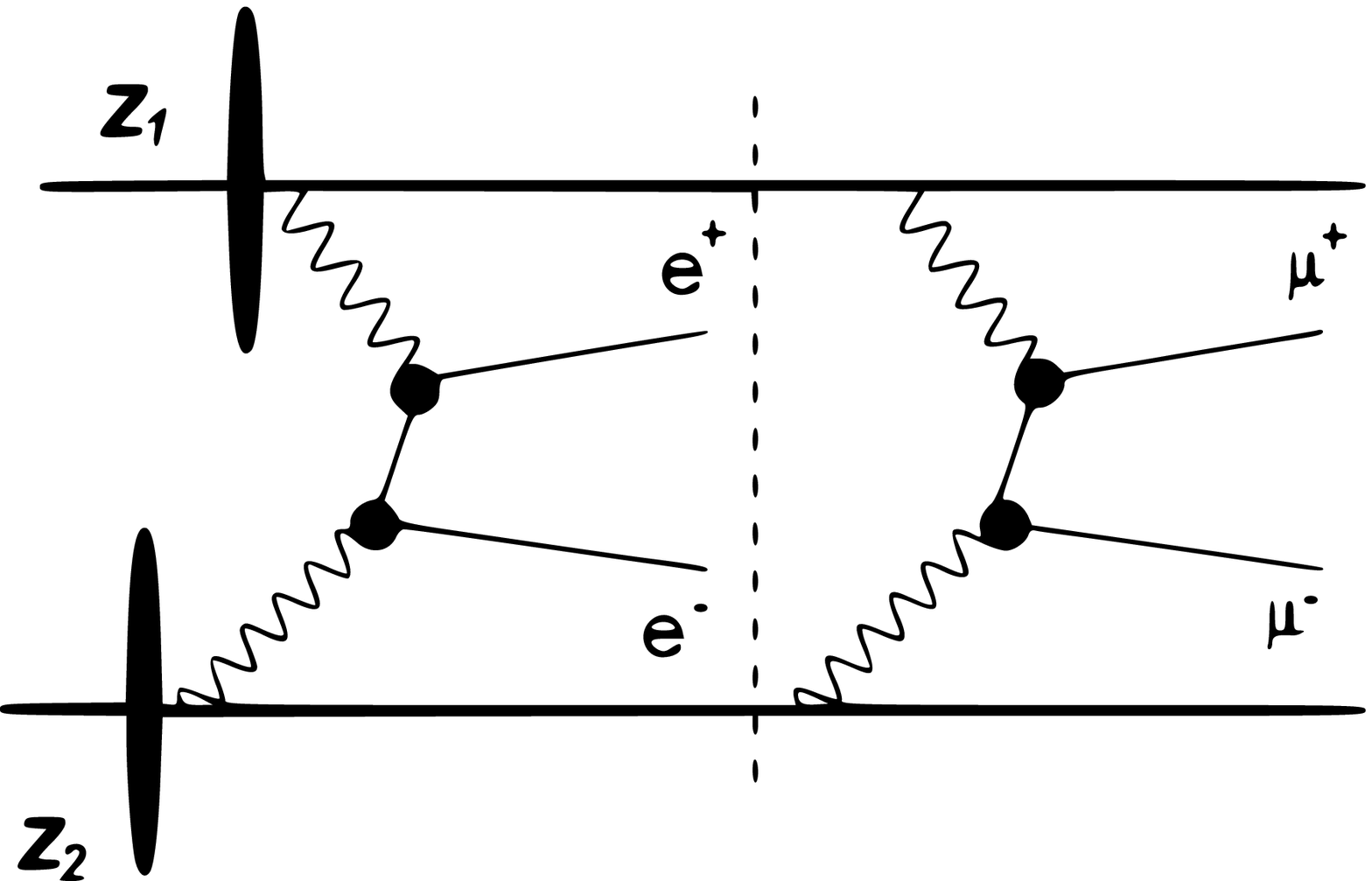}}\,\,\,\,\,\,\,\,\,                
  \subfloat[]{\includegraphics[width=7.0cm,height=4.0cm]{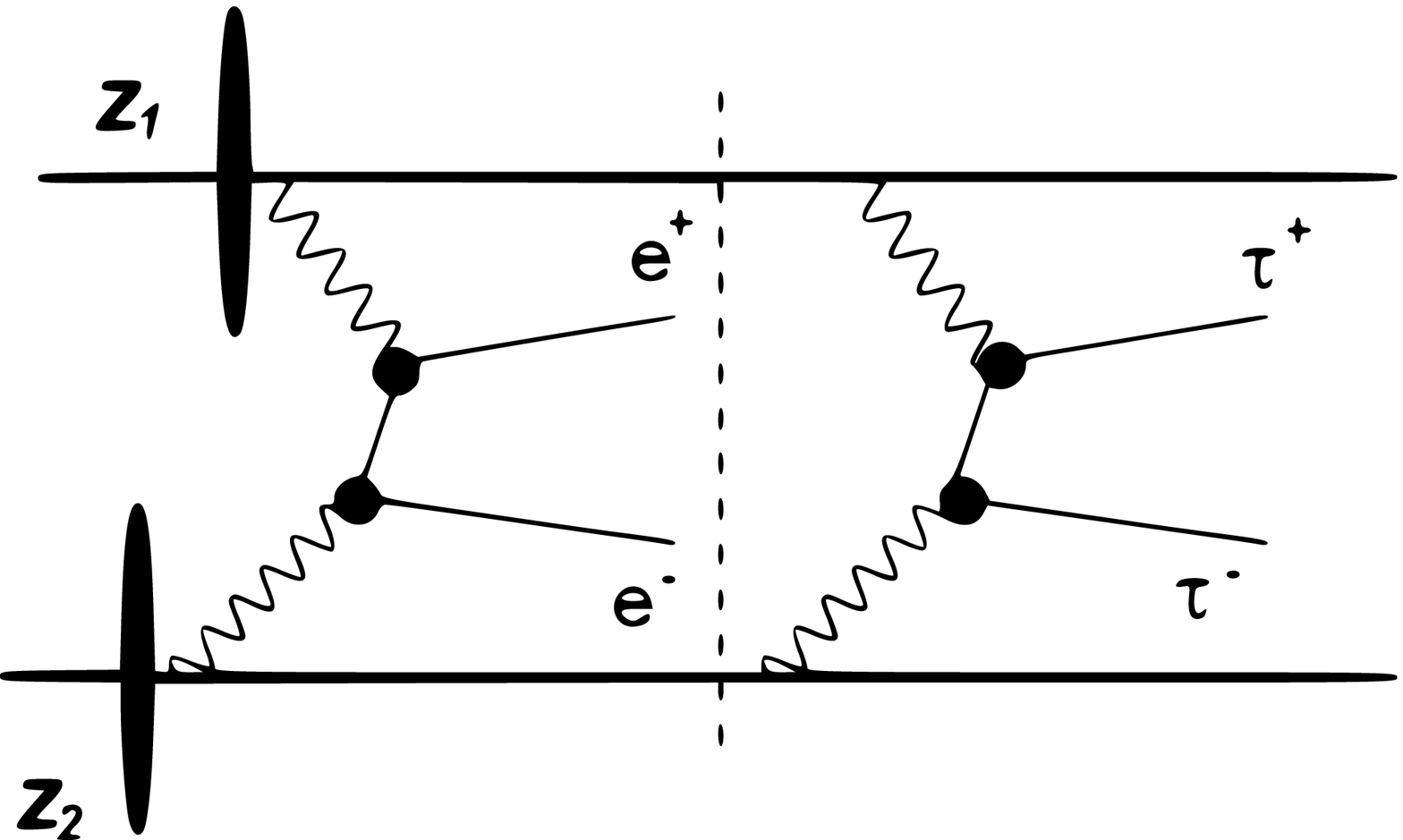}}\\
  \subfloat[]{\includegraphics[width=7.0cm,height=4.0cm]{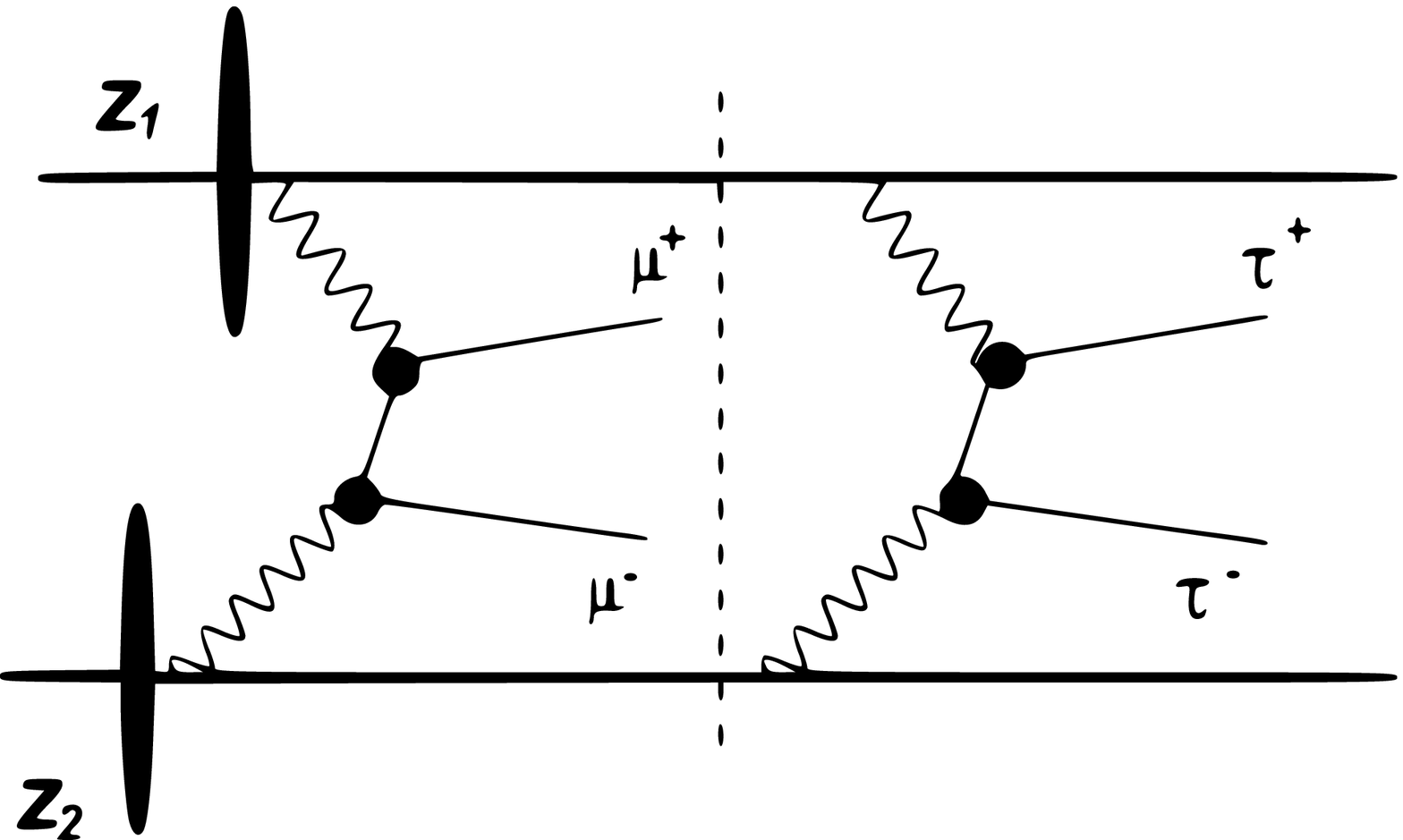}}\,\,\,\,\,\,\,\,\,                
  \subfloat[]{\includegraphics[width=9.0cm,height=4.0cm]{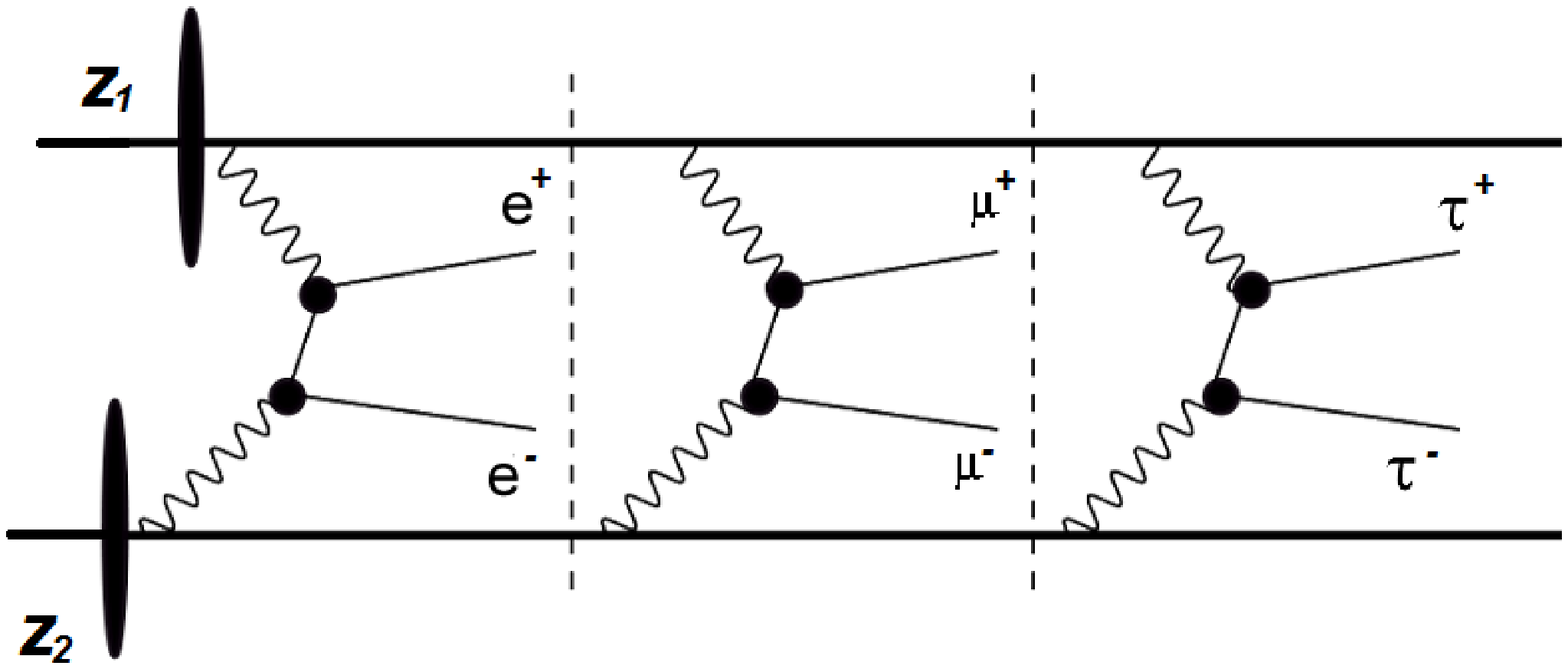}}
  \caption{Second order Feynman diagrams for producing simultaneously (a) electron and muon pair production  (b) electron and tauon pair production
 (c) muon and tauon pair production and  (d) electron, muon and tauon pair production.}
\label{fig2}
\end{figure}
\begin{figure}
\includegraphics[width=8.0cm,height=6.0cm]{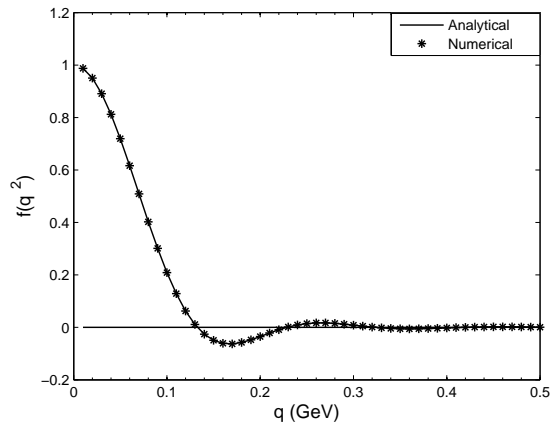} 
\caption{The electromagnetic form factor for gold. The solid line is the analytic form of Woods-Saxon distribution and the star line is the numerical calculation of the same equation.}
\label{fig3}  
\end{figure}

Including all these equations and definitions, we can write the total cross section
of free lepton pair production cross section as
\begin{equation}
\sigma=\int d^{2}b\sum_{k>0} \sum_{q<0} \mid <\chi^{+}_{k}\mid S \mid \chi^{-}_{q}> \mid ^{2}
\label{sigma}
\end{equation}
where S is scattering matrix, $\mid \chi^{+}_{k} >$ and $\mid \chi^{-}_{q} >$ are single-particle and single anti-particle states respectively. By
following the steps used in references \cite{bs,gwuse,guclu}, lepton pair production cross section can be written as

\begin{eqnarray}
\label{tot1}
\sigma=\dfrac{1}{4\beta^{2}}\sum_{\sigma_{k}}\sum_{\sigma_{q}}\int \dfrac{dk_{z}dq_{z}d^{2}p_{\perp}}{(2\pi)^{4}} \int \dfrac{d^{2}k_{1\perp} d^{2}q_{2\perp}}{(2\pi)^{4}}
\times\mid F(k_{\perp}-p_{\perp};w_{1})F(k_{\perp}-p_{\perp};w_{2})T(p_{\perp}:+\beta) \\ \nonumber
+F(k_{\perp}-p_{\perp};w_{2})F(k_{\perp}-p_{\perp};w_{1})T(p_{\perp}:-\beta)\mid^{2}.
\end{eqnarray}
The frequencies of $\omega_{1}$ and $\omega_{2}$ are from the heavy ions 1 and 2, respectively.
The summation over $k$ is valid for the positive energy states, and the summation over $q$ is limited to the negative energy states in the Dirac sea.
$k_{z}$ and $q_{z}$ are the longitudinal momentums and $k_{\perp}$ and $q_{\perp}$ are transverse momentums of the produced leptons and anti-leptons, respectively. 
In addition to this, $p_{\perp}$ is the transverse momentum of the intermediate states, however longitudinal momentum of the intermediate states $p_{z}$ are the fixed by the 
momentum conservation. Finally, we can write the reduced Feynman amplitude $T(p_{\perp};\beta)$ as
\begin{eqnarray}
T(p_{\perp};\beta)=\sum_{s}\sum_{\sigma_{p}}(E_{p}^{(s)}-(E_{k}^{(+)}/2 + \beta(k_{z}-q_{z})/2))^{-1}
\times<u_{\sigma_{k}}^{(+)}\mid(1-\beta\alpha_{z})\mid u_{\sigma_{p}}^{(s)}><u_{\sigma_{p}}^{(s)}\mid(1+\beta\alpha_{z})\mid u_{\sigma_{q}}^{(-)}>
\label{t1}
\end{eqnarray}
and $F(q:w)$ is the scalar part of the field from the heavy ions
\begin{eqnarray}
F(q:w)=\dfrac{4\pi Z\gamma^{2}\beta^{2}}{w^{2}+\beta^{2}\gamma^{2}\mid q \mid^{2}}f_{Z}(q^{2})
\end{eqnarray}
where $f_{Z}(q^{2})$ is the nuclear form factor defined in Eq. \ref{ffactor}, and $u_{\sigma_{p}}^{(s)}$ is the spinor part
of the states $\chi^{(s)}$.

Equation \ref{tot1} gives us the total cross section integrated over the impact parameters. However, if the impact parameter dependence cross section is
needed, then we should keep the impact parameter and integrate the cross section equation over all variables. When we do this, we can obtain  the 
equation in the form of 
\begin{equation}
\dfrac{d\sigma}{db}=\int^{\infty}_{0}dq qbJ_{0}(qb)\mathcal{F}(q)
\label{dsigma}
\end{equation}
where the equation includes zeroth order Bessel function. This Bessel function oscillates rapidly for large $b$ values, therefore  Monte Carlo Techniques 
can not be applied here. To solve the integral, the integration is divided into parts and calculated separately
\begin{eqnarray}
\mathcal{F}(q) = \dfrac{\pi}{8\beta^{2}}\sum_{\sigma_{k}}\sum_{\sigma_{q}}\int^{2\pi}_{0} d\phi_{q}\int \dfrac{dk_{z}dq_{z}d^{2}k_{\perp}d^{2}Kd^{2}Q}{(2\pi)^{10}}\\ \nonumber
\times [(F(\dfrac{Q-q}{2};w_{1})F(-K;w_{2})T(k_{\perp}-\dfrac{Q-q}{2};\beta)
+F(\dfrac{Q-q}{2};w_{1})F(-K;w_{2})T(k_{\perp}-K;-\beta))\\  \nonumber
\times (F(\dfrac{Q+q}{2};w_{1})F(-q-K;w_{2})T(k_{\perp}-\dfrac{Q+q}{2};\beta) 
+F(\dfrac{Q+q}{2};w_{1})F(-q-K;w_{2})T(k_{\perp}+q-K;-\beta))]
\end{eqnarray}
where $\mathcal{F}(q)$ function is calculated by Monte Carlo Technique for a fixed value of q. 
When the technique is implemented for Au+Au collision at 3400 and 100 GeV energies, the behaviour of the 
function is obtained as an exponential form. We have used sufficient Monte Carlo points to obtain adequate 
accuracy to obtain $F(q)$ function. A smooth function 
\begin{equation}
\mathcal{F}(q)= \mathcal{F}(0)e^{-aq}=\sigma_{T}e^{-aq}
\end{equation}
is then fit to the calculated values and obtain the $a$ values
which are tabulated in Table \ref{avalue}. $\mathcal{F}(0)$ correspond to the total cross section at $q=0.0$ value, and $a$ is slopes of the function.
After determining the $\mathcal{F}(q)$ function, we can insert it in the Eq. \ref{dsigma} 
\begin{eqnarray}
\dfrac{d\sigma}{db}=\mathcal{F}(0)\int^{\infty}_{0}dqqbJ_{0}(qb)e^{-aq}
=\sigma_{T}\dfrac{ab}{(a^{2}+b^{2})^{3/2}}
\end{eqnarray}
and obtain smooth and well behaved impact parameter dependence cross section expression. This equation is
valid for all kind of leptons. The value of $a$ and the total cross section $\sigma_{T}$ are different
for electrons, muons and tauons and they are tabulated in Tables \ref{avalue} and \ref{tot}.
\begin{table}[ht]
	\caption{a values are obtained for electron, muon, tauon pairs at 100 GeV and 3400 GeV. $\lambda^{e}_{c}$, $\lambda^{\mu}_{c}$ and $\lambda^{\tau}_{c}$ are Compton wavelength for electron,muon.tauon respectively.} 
	\centering  
	\begin{tabular}{c c c } 
		\hline\hline   \\                     
		Lepton Pairs\,\,\,\,\,\,  & \,\,\, $100$ (GeV)\,\,\, &\,\,\,\, $3400$ (GeV) \,\,\,\, \\ [0.5ex] 
		\hline  \\                
		$e^{\pm}$ & 5.30 \, $\lambda^{e}_{c}$ & 12.55 \,\, $\lambda^{e}_{c}$   \\ 
		$\mu^{\pm}$ &   13.85  $\lambda^{\mu}_{c}$  & 19.24 \,\,\,\,$\lambda^{\mu}_{c}$  \\
		$\tau^{\pm}$ &  84 \,\,\,\,\,\,   
		$\lambda^{\tau}_{c}$ & 187.57 $\lambda^{\tau}_{c}$   \\ [1ex]      
		\hline  
	\end{tabular}
	\label{avalue} 
\end{table}
\begin{table}[ht]
	\caption{Total Cross Section values are shown lepton pairs at RHIC and LHC. First row is shown producing electron, muon, tauon pairs separately while second row is shown producing two different type lepton pairs at same time. Third row is shown producing electron-muon-tauon pairs at same time. } 
	\centering  
	\begin{tabular}{c c c } 
		\hline\hline   \\                     
		Lepton Pairs\,\,\,\,\,\,  & \,\,\, Total Cross Section at RHIC\,\,\, &\,\,\,\, Total Cross Section at LHC \,\,\,\, \\ & (barn) & (barn)  \\ [0.5ex] 
		
		\hline  \\                
		$e^{\pm}$ &  $34566\,\,\,\,\,\,\,\,\,\,\,\,\,$     &  $210254 \,\,\,\,\,\,\,\,\,\,\,$     \\ 
		$\mu^{\pm}$ &     $0.2\,\,\,\,\,\,\,\,\,\,\,\,\,\,\,\,\,\,\,\,\,$     & $2.3\,\,\,\,\,\,\,\,\,\,\,\,\,\,\,\,\,\,\,\,\,\,$   \\
		$\tau^{\pm}$ &   $3.7\times10^{-6}\,\,$  & $1.4\times10^{-3}\,\,$    \\ [1ex]      
		\hline \\
		$e^{\pm}$-$\mu^{\pm}$ &  $0.039\,\,\,\,\,\,\,\,\,\,\,\,\,\,\,$  &  $0.22\,\,\,\,\,\,\,\,\,\,\,\,\,\,\,\,\,\,$  \\ 
		$e^{\pm}$-$\tau^{\pm}$ &     $7.3\times10^{-7}\,\,$  &  $9.4\times10^{-5}\,\,$  \\
		$\mu^{\pm}$-$\tau^{\pm}$ &  $9.6\times10^{-9}\,\,$   
		&  $1.2\times10^{-5}\,\,$  \\ [1ex]      
		\hline \\
		$e^{\pm}$-	$\mu^{\pm}$-	$\tau^{\pm}$ &$1.92\times10^{-9}\,$ & $1.1\times10^{-6}\,\,$   \\ 
		\\ [1ex]      
		
		\hline  
	\end{tabular}
	\label{tot} 
\end{table}

The lowest order perturbative result of probability can be written as
\begin{eqnarray}
\mathcal{P}(b)=\sum_{k>0}\sum_{q<0}\mid <X_{k}^{+} \mid S \mid X^{-}_{q}> \mid^{2}
\end{eqnarray}
which is equal to
\begin{equation}
\mathcal{P}(b)=\dfrac{1}{2\pi b}\dfrac{d\sigma}{db}
=\sigma_{T}\dfrac{a}{2\pi(a^{2}+b^{2})^{3/2}}.
\end{equation}
However, for small impact parameters and high energies unitarity is violated so the probability becomes
greater than one. Therefore we can use the Poisson distribution \cite{glues,ahtb,guclu} whose mean value is $\mathcal{P}(b)$,
\begin{equation}
P_{N}(b)=\dfrac{\mathcal{P}(b)^{N}exp(-\mathcal{P}(b))}{N!}
\end{equation}
where N is number of pairs. In order to obtain 1-pair cross section (N = 1) $\sigma_{1pair}$, we simply integrate 
the 1-pair probability over the impact parameter $b$
\begin{equation}
\sigma_{1pair}=\int d^{2}bP_{1}(b).
\end{equation}
Our main task is to calculate the probabilities of more than one type of leptons simultaneously. 
For this, we can use the factorization method \cite{bhatk,bkn,bgkn} to calculate $e\mu, e\tau, \mu \tau$ pair production probabilities as
\begin{equation}
P_{1}^{l_{1}l_{2}}(b)=\{\,P_{1}^{e}(b)P_{1}^{\mu}(b),\,P_{1}^{e}(b)P_{1}^{\tau}(b),\,P_{1}^{\mu}(b)P_{1}^{\tau}(b)\,\}
\end{equation}
where $l_{1},l_{2} =e,\mu,\tau$ and these probabilities are plotted in Fig. \ref{fig5}.
Similarly, an impact parameter dependence probability of producing for three different types of leptons
can be written as 
\begin{equation}
P_{1}^{e \mu \tau}(b)=P_{1}^{e}(b)P_{1}^{\mu}(b)P_{1}^{\tau}(b)
\end{equation}
it is plotted in Fig. \ref{fig6}. For the total cross sections we have to integrate above probabilities over the the impact
parameter $b$
\begin{eqnarray}
\sigma_{T}^{l_{1}l_{2}}=\int^{\infty}_{0}{2\pi b}{db}P_{1}^{l_{1}}(b)P_{1}^{l_{2}}(b)
\end{eqnarray}
and for the total cross section of the three type of lepton pair production
\begin{eqnarray}
\sigma_{T}^{e \mu \tau}=\int^{\infty}_{0}{2\pi b}{db}P_{1}^{e}(b)P_{1}^{\mu}(b)P_{1}^{\tau}(b).
\end{eqnarray}
All these calculated total cross sections are tabulated in Table \ref{tot}.
\section{\label{sec:level5}Result}\label{s5}

In this work, we study the total cross sections, the impact parameter dependent cross sections and the probabilities of lepton 
pairs which are produced by Au+Au collisions at the LHC and the RHIC energies. We especially focus on the probabilities of 
different types lepton pairs that are created simultaneously.

Since the Compton wavelength of the electron is much larger than the radius of the colliding heavy ions, 
the form factors of nucleus almost have no effect for the production of electron-positron pairs.
On the other hand, the Compton wavelengths of muons and tauons are smaller than the radius of the colliding 
heavy ions, therefore the effects of the form factors become dominant for the pair production. At the RHIC and the LHC 
energies, the muon pair production is reduced by about 3 and 2 factors, respectively. On the other hand, at the RHIC 
and the LHC energies, the tauon pair production is reduced by about 100 and 5 factors, respectively.

\begin{figure}[h]
  \centering
  \subfloat[]{\includegraphics[width=7.0cm,height=7.0cm]{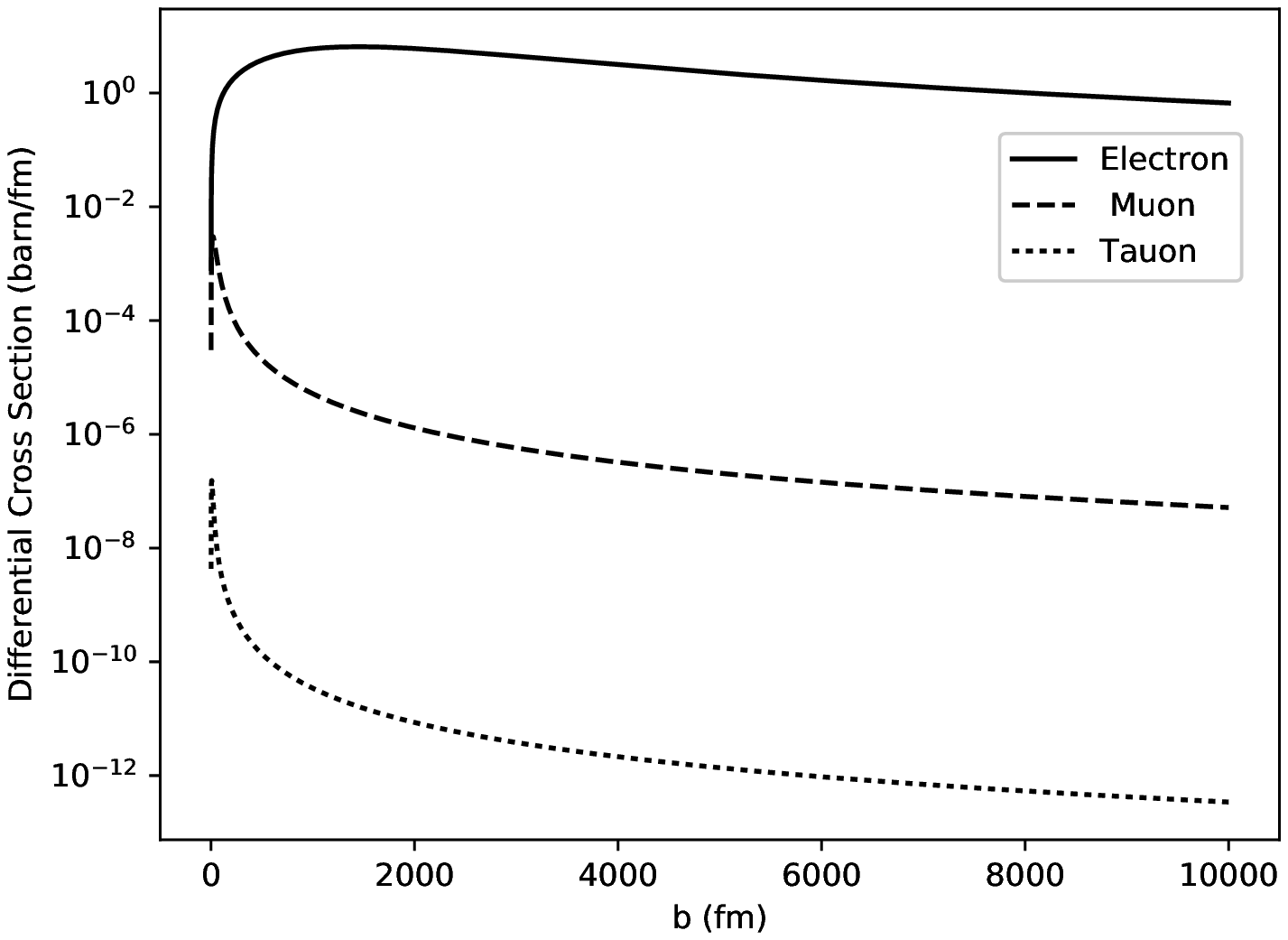}}                
  \subfloat[]{\includegraphics[width=7.0cm,height=7.0cm]{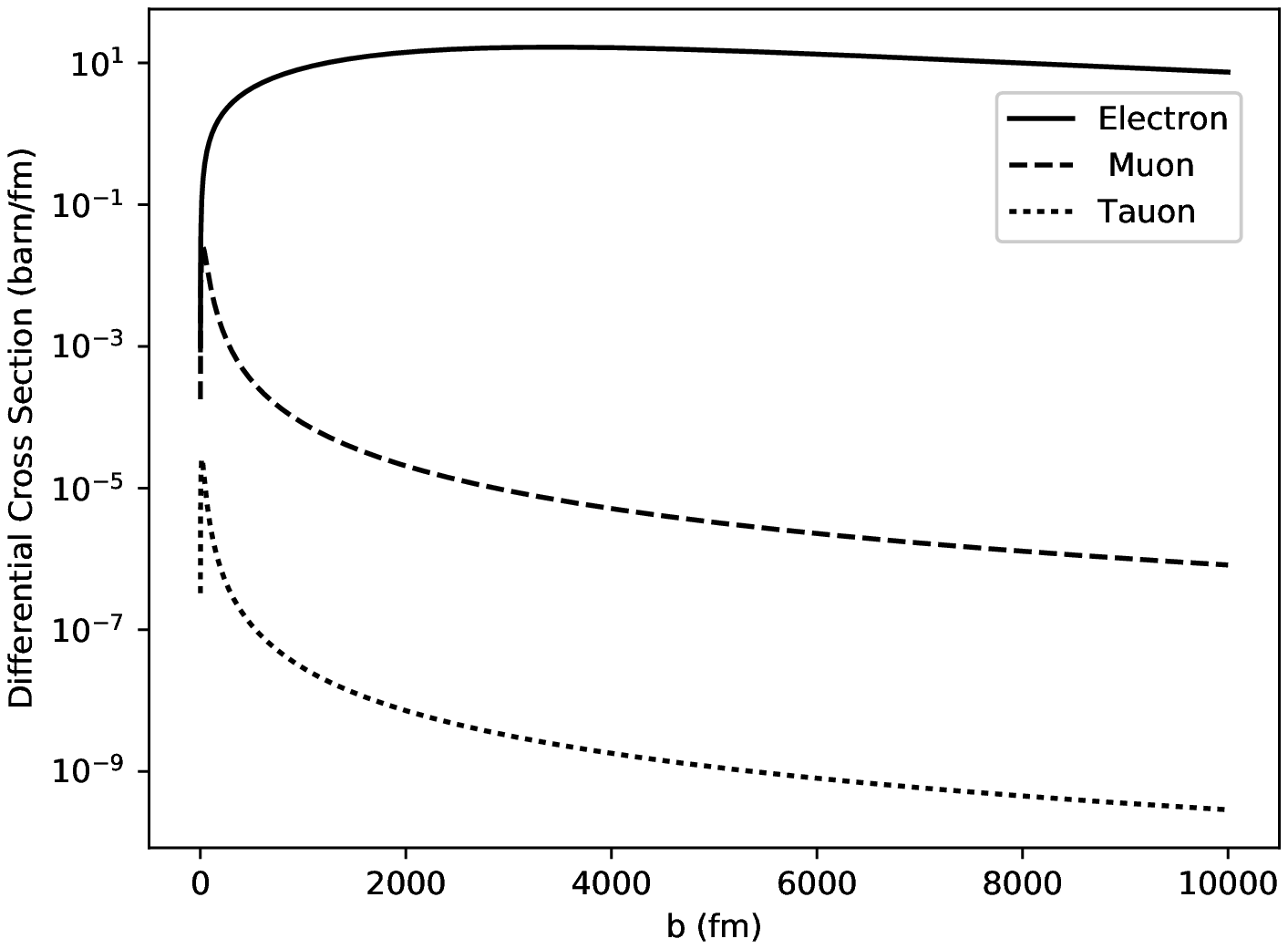}}
  \caption{(a) Impact parameter dependent cross section is indicated for lepton pairs which are produced by Au+Au ion collisions at 100 GeV energy value.  (b) Impact parameter dependent cross section is indicated for lepton pairs which are produced by Au+Au ion collisions at 3400 GeV energy value. }
 
\label{fig4}
\end{figure}
The cross section calculations involve nine dimensional integrals and they can not be calculated analytically, 
so we have used Monte-Carlo Method to calculate the equations numerically. We use two millions and more 
Monte-Carlo points and we have reached to the sufficient convergence to obtain the total cross section of electron, muon, and tauon pairs. 

In order to calculate the impact parameter dependent cross section, we firstly obtained values of $a$ for each type of lepton.
The values of $a$ are tabulated in table \ref{avalue} for all leptons. Once we determine
this parameter all calculations can be  easily done. In table \ref{tot}, we have shown the total cross sections of producing
for all combinations of the lepton pairs. In the first row of the table, we can see the single pair production 
cross sections. This clearly shows that when the mass of lepton becomes heavier, the cross section decreases rapidly.
On the other hand, in the second row, we have shown the simultaneous production of total cross sections for all two combinations of leptons.
These results show that at the LHC energies the cross sections are 2 or 4 order of magnitude greater than the RHIC energies.
Finally at the third row, the simultaneous total cross sections of all three type lepton pairs are written.
At the LHC energies, the observing this production is nearly three order of magnitude greater than the RHIC energies.

The impact parameter dependent cross sections of leptons are drawn in Fig.\ref{fig4} (a) for the RHIC and (b) for the LHC energies.
The behaviour of the both figures are nearly the same, however the impact parameter dependence 
is enhanced by a factor of one or two order of magnitude at the LHC energies. In Table \ref{table1} and Table \ref{table2}, it is 
clear that the relativistic heavy ions are accompanied by the strong electromagnetic fields. In the Weizs\"{a}cker-Williams approach, 
the combined fields of electric and magnetic fields from the ions may be treated as a flux of nearly-real virtual photons.
For two ions separated by a large $b$, the most favourable pair production point is midway between the two ions.  
At this point, the maximum photon energy is $k_{max}=2 \gamma \hbar c/b$, where $\gamma$ is the Lorentz boost of the ion, 
and $b$ is the impact parameter which is the transverse distance from the ion. At the RHIC, for $b=2000 fm$, $k_{max}= 20 MeV$
and at the LHC,  $k_{max}= 680 MeV$.  For the production of $\tau$ pairs, the impact parameter dependence cross section increased almost by 1000 times from the RHIC to the LHC
energies at the impact parameter 2000 fm, however the ratio of $k_{max}$ increased by $680/20 = 34$, only. 
To calculate the percentage of the contribution to the total cross section from greater than 2000 fm, we can use the following
integration
\begin{eqnarray}
(\int^{\infty}_{2000fm}\dfrac{d\sigma}{db}db)/(\int^{\infty}_{0}\dfrac{d\sigma}{db}db)
\end{eqnarray}
and we can show that the contribution to the total cross section of impact parameters greater than 2000 fm is nearly 0.5 \% at the RHIC and 1 \% at the LHC
for the $\tau$ productions.
This shows that the differential cross sections decrease fast for the large impact parameters. For the similar calculations
are done for all type of leptons and the results are tabulated in Table \ref{partial}. It is interesting that,
for the electron-positron pair production, $72 \%$ of the total cross section comes from the impact parameters which are greater than $2000 fm$ 
at RHIC and $92 \%$ at LHC. 
 
\begin{figure}[h]
  \centering
  \subfloat[]{\includegraphics[width=7.0cm,height=7.0cm]{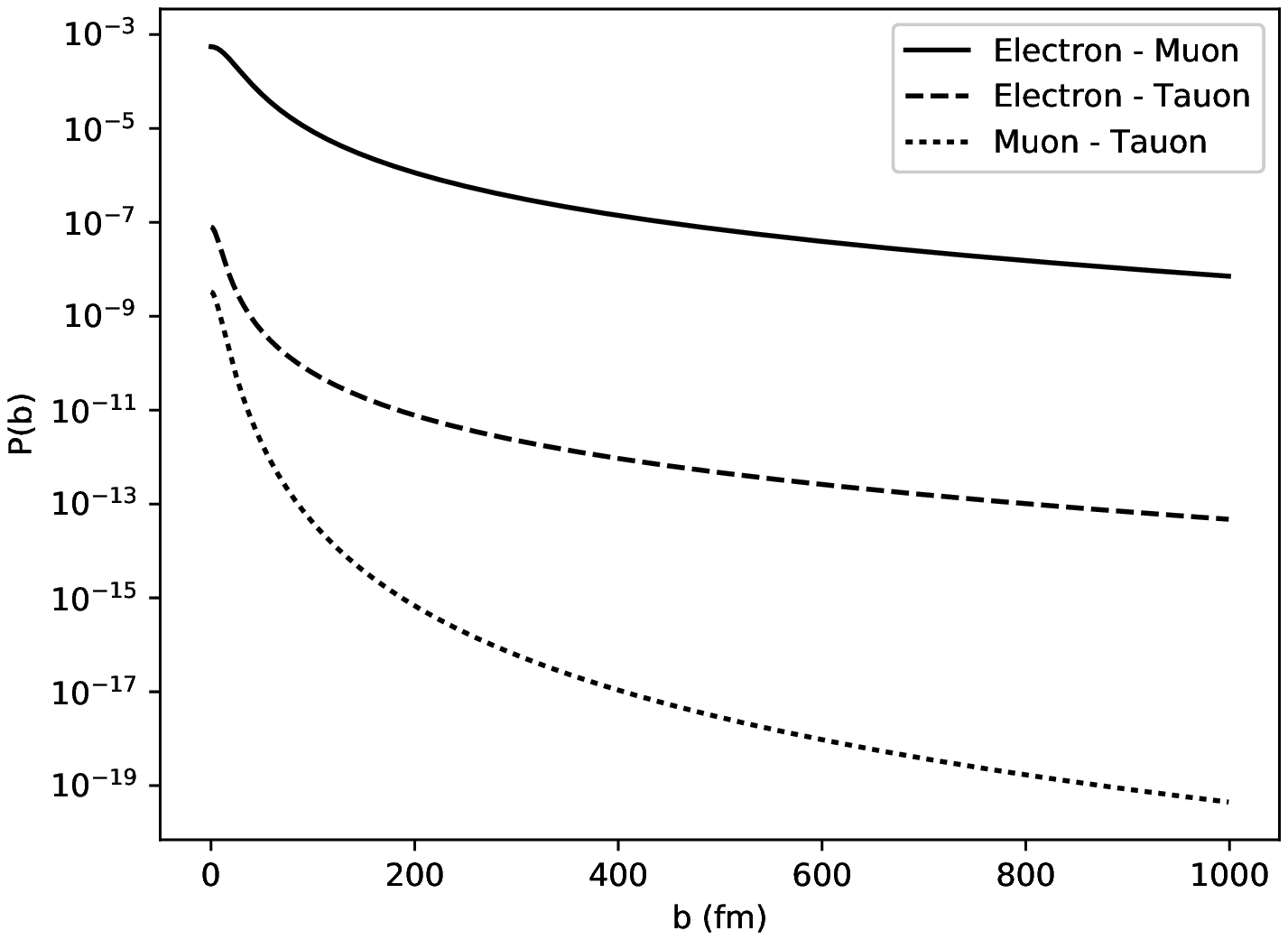}}                
  \subfloat[]{\includegraphics[width=7.0cm,height=7.0cm]{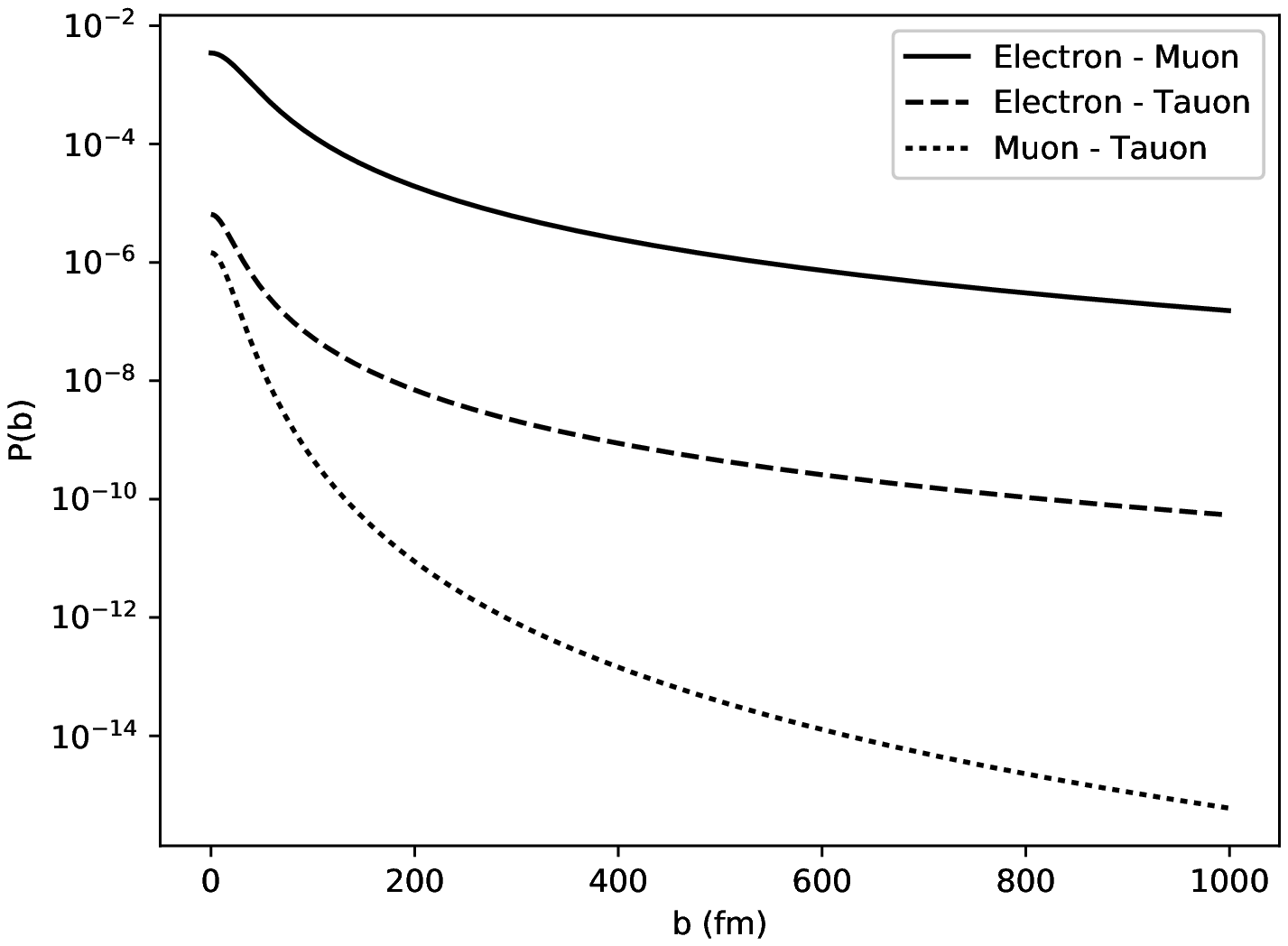}}
  \caption{ (a) The behaviours of the probability values are shown depending on impact parameter 'b' for two of $e^{\pm}-\mu^{\pm}-\tau^{\pm}$ pairs which are produced by Au+Au collisions at 100 GeV energy value at same time.  (b) The behaviours of the probability values are shown depending on impact parameter 'b' for two of $e^{\pm}-\mu^{\pm}-\tau^{\pm}$ pairs which are produced by Au+Au collisions at 3400 GeV energy value at same time.
 }
\label{fig5}
\end{figure}

In Fig. \ref{fig5}, we have plotted the impact parameter dependence probabilities of producing two different lepton
pairs simultaneously. Again in these figures, the patterns are nearly the same for both the RHIC and the LHC energies.
However, since the masses of muon and tauon are much larger than the electrons, the probabilities of producing 
these heavy leptons are much smaller than involving the electrons. The clear enhancement at the LHC energies is
also shown in the these figures.  

\begin{figure}
	\includegraphics[width=7.0cm,height=7.0cm]{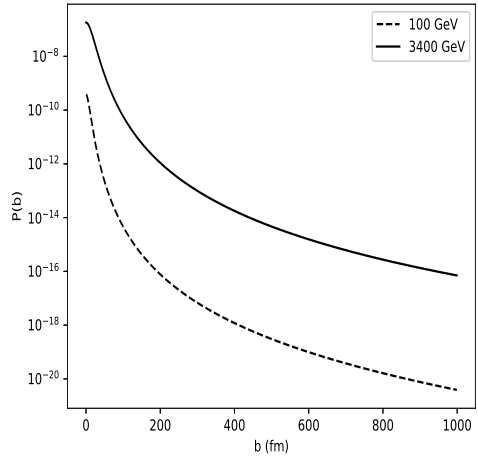} 
	\caption{The probability of lepton pairs which are produced by Au+Au collisions at 100 GeV and 3400 GeV energy values at same time is shown depending on impact parameter  'b'. }
	\label{fig6} 
\end{figure}

Finally in Fig. \ref{fig6}, we can see the impact parameter dependence for the probabilities of simultaneous production of all these three 
leptons for the RHIC and the LHC energies. This plot shows that the simultaneously observing the three types leptons increase 4 or 5 order of
magnitude at the LHC when it is compared with the RHIC.

Our results are also in agreement with  Ref.\cite{glues} for the electron-positron pair production. On the other hand, the results in
Ref. \cite{bhatk}, the muon pair production cross sections are slightly greater than our results. The reason for that, it could
be the choice of the nuclear form factors where they have used the monopole form factor however, we have used the  Wood-Saxon form factor in the equations.

Recent calculation \cite{knsbg} shows the results of all kinds of lepton pair productions from the relativistic
heavy ion collisions by using the Monte Carlo Program which is called STARlight. It is possible to compare our results
with the results that comes out from this program in future.
The heavy leptons and more than one type of leptons are produced in a strong field of ultra-relativistic heavy ions.
As we calculated the cross sections of these productions, it is clear that the cross sections are in the measurable ranges at the LHC.
As it is pointed out in Ref. \cite{baur}, because of very low transverse momentum $p_{\perp}$, it could be difficult to measure
these produced leptons. For the next work, it is our task to calculate numerically the transverse momentums of the leptons
and compare them with the longitudinal momentums.
\begin{table}[ht]
	\caption{The percentages of partial total cross sections of lepton pairs integrated between 2000 fm and $\infty$ } 
	\centering  
	\begin{tabular}{c c c } 
		\hline\hline   \\                     
			Lepton Pairs\,\,\,\,\,\,  & at RHIC\,\,\, & at LHC \,\,\,\, \\ & ( \%) & (\%)  \\ [0.5ex] 
		
		\hline  \\                
		$e^{\pm}$    & $72$    & $92$      \\ 
		$\mu^{\pm}$  & $1$    & $2$   \\
		$\tau^{\pm}$ & $0.5$     & $1$    \\ [1ex]      

		\hline  
	\end{tabular}
	\label{partial} 
\end{table}

\section{ACKNOWLEDGMENTS}
This research is partially supported by the Istanbul Technical University. We personally thank S. R. Klein for valuable advice in calculating
the cross sections and for the careful reading of our article.
%

%
\end{document}